\documentclass[twocolumn]{aa}
\usepackage{graphicx}
\usepackage{txfonts}
\usepackage{natbib}
\bibpunct{(}{)}{;}{a}{}{,}
\usepackage{color}
\definecolor{red}{rgb}{0.7,0,0}
\definecolor{blue}{rgb}{0,0,0.7}
\begin{document}
   \title{First INTEGRAL observations of GRS~1915$+$105\thanks{Based on observations with {\it INTEGRAL}, an ESA project with instruments and science data center funded by ESA and member states (especially the PI countries: Denmark, France, Germany, Italy, Switzerland, and Spain), the Czech Republic, and Poland and with the participation of Russia and the US.}}


   \author{D.C. Hannikainen
          \inst{1} \and
          O. Vilhu\inst{1,2} \and J. Rodriguez\inst{3,2} \and 
          S. Brandt\inst{4} \and N.J. Westergaard\inst{4}
          \and N. Lund\inst{4} \and I. Moc{\oe}ur\inst{3} \and
          Ph. Durouchoux\inst{3} \and T. Belloni\inst{5} 
          \and A. Castro-Tirado\inst{6} \and
          P.A. Charles \inst{7} \and A.J. Dean\inst{7} \and
          R.P. Fender\inst{8} \and M. Feroci\inst{9} \and
          P. Hakala\inst{1} \and R.W. Hunstead\inst{10} \and
          C.R. Kaiser\inst{7} \and A. King\inst{11} \and
          I.F. Mirabel\inst{3} \and G.G. Pooley\inst{12} \and
          J. Poutanen\inst{13} \and K. Wu\inst{14} \and
          A.A. Zdziarski\inst{15} 
          }

   \offprints{D. Hannikainen: diana@astro.helsinki.fi}

   \institute{Observatory, PO Box 14, FIN-00014 University of Helsinki, Finland 
	 \and
              INTEGRAL Science Data Center, Chemin d'\'Ecogia 16, CH-1290 Versoix, Switzerland
         \and
              Centre d'Etudes de Saclay, DAPNIA/Service
              d'Astrophysique (CNRS FRE 2591), 
              Bat. 709, Orme des Merisiers, Gif-sur-Yvette Cedex 91191, France 
         \and
             Danish Space Research Institute, Juliane Maries Vej 30, Copenhagen O, DK-2100 Denmark
	 \and 
             INAF - Osservatorio Astronomico di Brera, via E. Bianchi 46, 23807 Merate (LC), Italy 
	 \and
	     Instituto de Astrof\'{\i}sica de Andaluc\'{\i}a (IAA-CSIC), PO Box 03004, 18080 Granada, Spain 
	 \and
	     Dept. of Physics and Astronomy, University of Southampton, Southampton SO17 1BJ, UK
	 \and 
             Astronomical Institute, ``Anton Pannekoek'', University of Amsterdam, Kruislaan 403, 1098 SJ Amsterdam, Netherlands
	 \and
	     Istituto Astrofisica Spaziale e Fisica Cosmica, Sezione di Roma, CNR, via Fosso del Cavaliere, 00133 Roma, Italy
	 \and
             School of Physics, University of Sydney, NSW 2006, Australia
	 \and
	     Theoretical Astrophysics Group, University of Leicester, Leicester LE1 7RH, UK
         \and
            Astrophysics Group, Cavendish Laboratory, University of Cambridge, Madingley Road, Cambridge CB3 0HE, UK	  
	 \and
            Astronomy Division, P.O.Box 3000, FIN-90014 University of Oulu, Finland
         \and
            MSSL, University College London, Holmbury St. Mary, Surrey, RH5 6NT, UK
         \and
            Nicolaus Copernicus Astronomical Center, Bartycka 18, 00-716 Warszawa, Poland
           }

   \date{Received ; accepted}

   \abstract{We present data from the first of six monitoring 
                Open Time observations 
		of GRS~1915$+$105 undertaken with the orbiting 
		INTEGRAL satellite. 
             The source was clearly detected with all three X-ray and 
                 gamma-ray instruments on board.
             GRS~1915$+$105 was in a highly variable state,
                 as demonstrated by the JEM~X-2 and ISGRI lightcurves. 
These  and simultaneous RXTE/PCA lightcurves point to a novel type of variability 
pattern in the source. In addition, we fit the combined JEM~X-2 and ISGRI spectrum between 
3--300~keV with a disk blackbody $+$ powerlaw model leading to typical parameter values found earlier  
at  similar luminosity levels.
A new transient, IGR~J19140$+$098, was discovered during the present observation.
   }

   \maketitle
%

\section{Introduction}
GRS~1915$+$105 has been extensively observed at all wavelengths ever
  since its discovery. 
It was originally detected as a hard X-ray source with 
  the WATCH all-sky monitor on the GRANAT satellite \citep{castro} with
  a flux of 0.35~Crab in the 6--15~keV range (Castro-Tirado et al. 1994). 
Subsequent monitoring with BATSE on CGRO showed it to be the 
  most luminous hard X-ray source in the Galaxy \citep{paciesas}, with 
  L$_{\rm 20-100 keV}\sim3-6\times10^{38}$~erg~s$^{-1}$ 
  (for a distance of 12.5~kpc). 
Apparent superluminal ejections have been observed from 
  GRS~1915$+$105 on at least two occasions: the first
  time in 1994 with the VLA \citep{mirabel94} and the second time in
  1997 with MERLIN \cite{fender99}. 
Both times the true ejection velocity was calculated to be $>0.9c$. 
Following the ejections of 1997, Fender et al. (1999) give an 
  upper limit for the distance to the source of 11.2$\pm$0.8~kpc.
High optical absorption ($\geq 33$ magnitudes)
  towards GRS~1915$+$105 prevented the identification of the non-degenerate 
  companion until recently.
However, using the VLT, Greiner, Cuby \& McCaughrean (2001) 
  identified the mass-donating star to be of spectral type K-M III and 
  hence deduced the mass of the black hole to be $14\pm4{\rm M}_{\sun}$.
The Rossi X-ray Timing Explorer (RXTE) has observed GRS~1915$+$105
  since its launch in late 1995 and has shown the source to be highly
  variable (see e.g. Belloni et al. 2000).
The source was detected up to $\sim$~700~keV during OSSE observations
  (Zdziarski et al. 2001).
  
GRS~1915$+$105 is being observed extensively with the 
  European Space Agency's {\it 
  International Gamma-Ray Astrophysical Laboratory} (INTEGRAL, Winkler
  et al. 2003) as
  part of the Core Program and also within the framework of an Open
  Time monitoring campaign. 
Here we present the results of the first observation of the Open Time
  monitoring campaign which took place on 2003~Mar~6.
We show the lightcurves and the spectra from ISGRI and JEM-X, in 
  addition to a SPI image. 
We also show that the source underwent a new type of variability
  during this observation, which is seen in the JEM~X-2 lightcurve
  and confirmed by simultaneous RXTE observations.

\section{Observations}

\begin{figure}[h]
   \centering
   \includegraphics[angle=0,width=9cm]{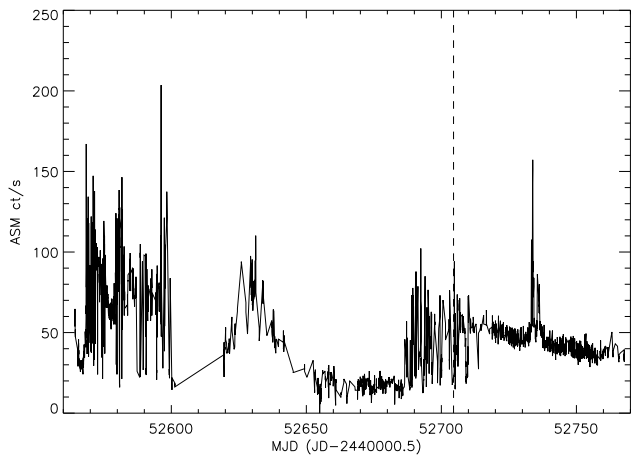}
   \caption{The RXTE/ASM 1.2--12~keV single dwell lightcurve. The starting
            date in the figure is 2002 Oct 17, INTEGRAL's launch
            date. The dashed line shows 2003 March 6, i.e.
            the date of our first observation. 75 ASM ct/s
            corresponds to 1 Crab. }
              \label{asm}
    \end{figure}

As part of a monitoring program which consists of six 
  100~ks observations separated by approximately one month, 
  INTEGRAL observed GRS~1915$+$105 for the first time for this
  campaign on 2003 Mar 6 beginning at 03:22:33UT.
Simultaneously with the INTEGRAL observations, we obtained
  pointed RXTE observations (Rodriguez et al., in preparation), 
  in addition to radio data from the Molonglo Observatory Synthesis
  Telescope, the Ryle Telescope and RATAN. 
These latter data will be dealt with in detail in a forthcoming
  paper which will cover all the observations (Hannikainen et al.,
  in preparation).
Figure~\ref{asm} shows the RXTE/ASM 1.2--12~keV lightcurve beginning with 
  INTEGRAL's launch on 2002 Oct 17. 
The dashed line indicates the date of the 100~ks INTEGRAL
  observation.
As 75 ASM ct/s corresponds to 1 Crab, our observation was undertaken
  while the source had a luminosity of $\sim$560~mCrab (in the
  1.2--12~keV ASM range), which is close to the average luminosity
  of the source. 

The INTEGRAL observations were undertaken using the hexagonal dither 
  pattern (Courvoisier et al. 2003): 
   this consists of a hexagonal pattern around the nominal target location 
   (1 source on-axis pointing, 6 off-source pointings, each 2 degrees apart
   and each science window of 2200 s duration).
This means that GRS~1915$+$105 was always in the field-of-view
   of all three X-ray (JEM~X-2) and gamma-ray (IBIS and SPI) instruments
   throughout the whole observation.
The source was clearly detected by all three X-ray and gamma-ray
  instruments as shown below (Figs.~\ref{image}, \ref{spi},
  \ref{jemx} and \ref{spectrum}).
(Due to high extinction towards the source, the target was not visible
   in the optical monitor.)
The properties of the INTEGRAL instruments are described in detail 
  elsewhere in this journal and so we give only a brief introduction. 
In this section we present the results of the INTEGRAL 
  observation of 2003~Mar~6. 
  
  All  data reductions were performed using standard software explained elsewhere
  in the present issue, using instrument responses available in June 2003.

\subsection{ISGRI}

  \begin{figure}[h]
   \centering
   \includegraphics[angle=0,width=9cm]{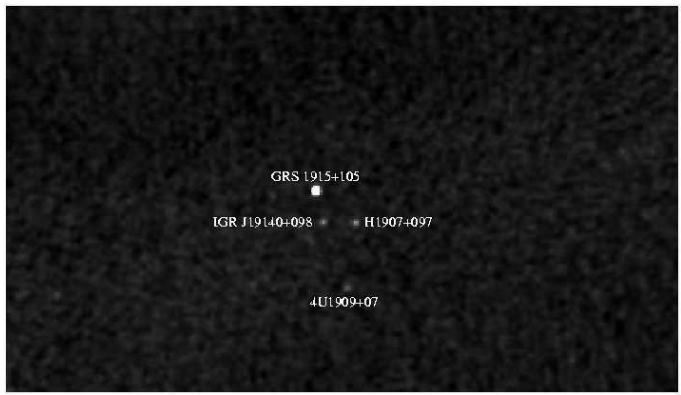}
   \caption{The IBIS/ISGRI 20--40~keV ($\sim22^{\circ}$ width and
   $\sim12^{\circ}$ height) image, showing the location of 
            GRS~1915$+$105 and three other bright sources in the
   field, including the new transient IGR~J19140$+$098
   discovered during this observation.
   North is up and East is to the left}
              \label{image}
    \end{figure}
 
The Imager on Board the INTEGRAL Spacecraft (IBIS, Ubertini et al. 2003) is a
 coded mask instrument 
  designed for high angular resolution (12 arcmin, but source location down to 1 arcmin) 
  imaging in the energy range from $\sim20$ keV to $\sim10$ MeV. 
Its total field of view is $\sim30\times30$ degrees for zero response 
  with a uniform sensitivity 
  within the central $\sim10\times10$ degrees. 
It has $19\times19$ degrees FWHM, i.e. up to 50\% response of the
  instrument. 
The INTEGRAL Soft Gamma-Ray Imager (ISGRI, Lebrun et al. 2003) is the top layer of the
 IBIS detection plane, 
  which consists of an array of 128$\times$128 Cadmium Telluride
  (CdTe) square  pixels  (4 mm $\times$ 4 mm each) and covers the
  energy range from ~20 keV to a few hundred keV. 
Data reduction has been performed following the standard method
  described in Goldwurm et al. (2003) extracting from each 2200s 
  pointing the source positions and count rates. 
Then a mosaic was produced from the summation of all pointings.


Figure~\ref{image} shows the IBIS/ISGRI 20--40~keV ($\sim 22^{\circ} \times
  12^{\circ}$) mosaicked image of the field of GRS~1915$+$105,
  with an exposure time of 98300~s.
A preliminary instrumental background correction has been performed
  (Terrier et al. 2003).
Two other known sources,  H1907$+$097 and 4U~1909$+$07, are visible in the
  field.

In addition, a new transient, IGR~J19140$+$098 (SIMBAD corrected name
  IGR~J19140+0951) was discovered during the observation 
  of 2003~Mar~6 (Hannikainen, Rodriguez \& Pottschmidt 2003).

\begin{figure}[h]
  \centering
  \includegraphics[angle=0,width=9cm]{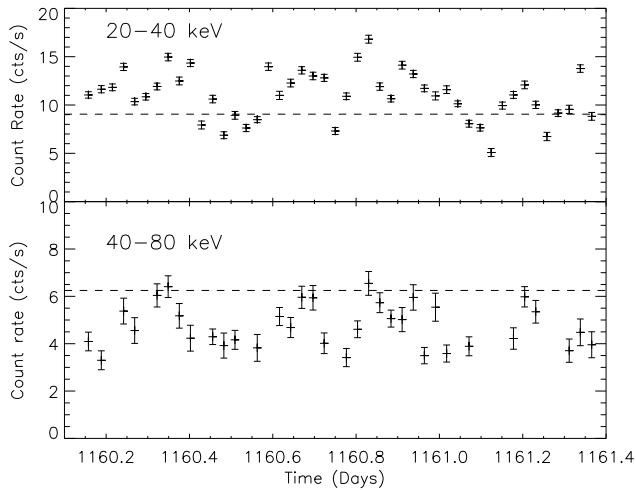}
  \caption{The ISGRI 20--40~keV ({\bf top}) and 40--80~keV ({\bf
  bottom}) lightcurves. The dashed horizontal lines indicate the 
  50~mCrab level in the given energy ranges showing, in
  particular how much steeper the source spectrum is compared to 
  the Crab. The bin size is 2200~s. The time of the observation is
  given in IJD, or ``ISDC Julian Date'', and is defined as the
  fractional number of days since 2000~Jan~1 (in other words
  IJD=JD$-$2,451,544.5).}
 \label{lcurve}
\end{figure}

Figure~\ref{lcurve} (top) shows the 20--40~keV and (bottom) the 
  40--80~keV ISGRI lightcurves. 
The horizontal line indicates the 50~mCrab level in the given
  energy ranges. 
As can be inferred, 
   the luminosity of the source varies  between $\sim30-100$~mCrab
   in the 20--40~keV range and between  $\sim24-50$~mCrab
   in the 40--80~keV range.

\subsection{SPI}

\begin{figure}[h]
\centering
 \includegraphics[angle=0,width=8cm]{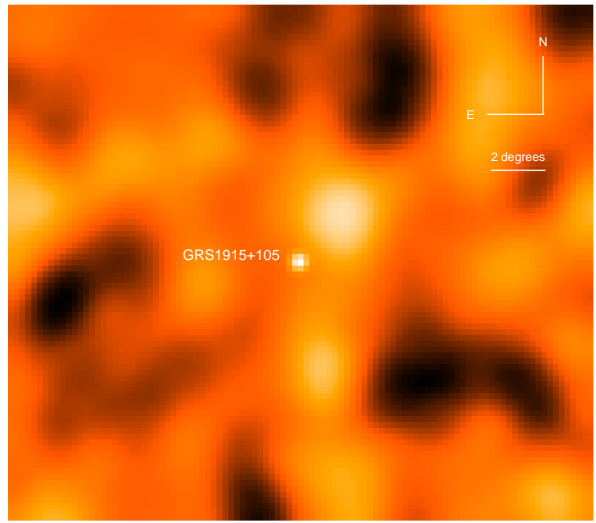}
\caption{A SPI 100--200~keV image of GRS~1915$+$105. 
     GRS~1915$+$105 
     is clearly visible at the center of
     the image (marked).
     The three other sources seen in Fig.~\ref{image} are much fainter in this energy range and not visible.   }
    \label{spi}
    \end{figure}

The Spectrometer on INTEGRAL (SPI, Vedrenne et al. 2003) is a coded mask instrument designed for high energy 
   resolution (2 keV at 1 MeV) spectroscopy of gamma-ray sources in the 20 keV to 8 MeV range. 
It consists of an array of 19 hexagonal high-purity Germanium detectors. 
The field of view is  16$^{\circ}$ in diameter with an angular resolution of 2$^{\circ}$. 
The resulting 100--200~keV SPI image is shown in Figure~\ref{spi}.
GRS~1915$+$105 is clearly detected up to 200~keV.

One point which must be noted is that the new transient,
  IGR~J19140+098, lies only 1.13$^{\circ}$ away from GRS~1915$+$105
  thus rendering the spectral extraction highly uncertain (due to
  source confusion) at this early stage of the mission.  

\subsection{JEM~X-2}

\begin{figure}[h]
\begin{tabular}{c}
 \includegraphics[angle=0,width=8.5cm]{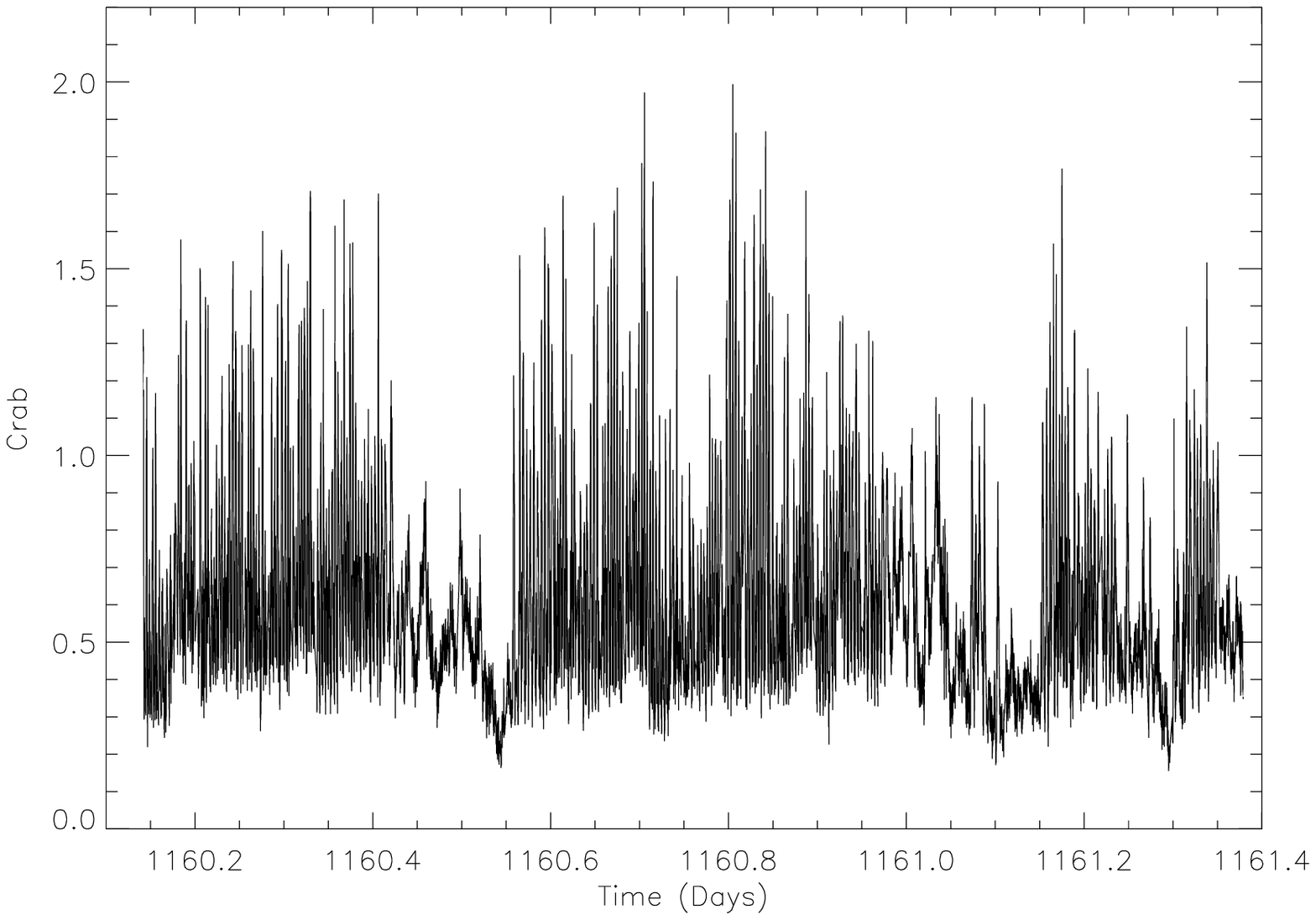} \\
 \includegraphics[angle=0,width=8.5cm]{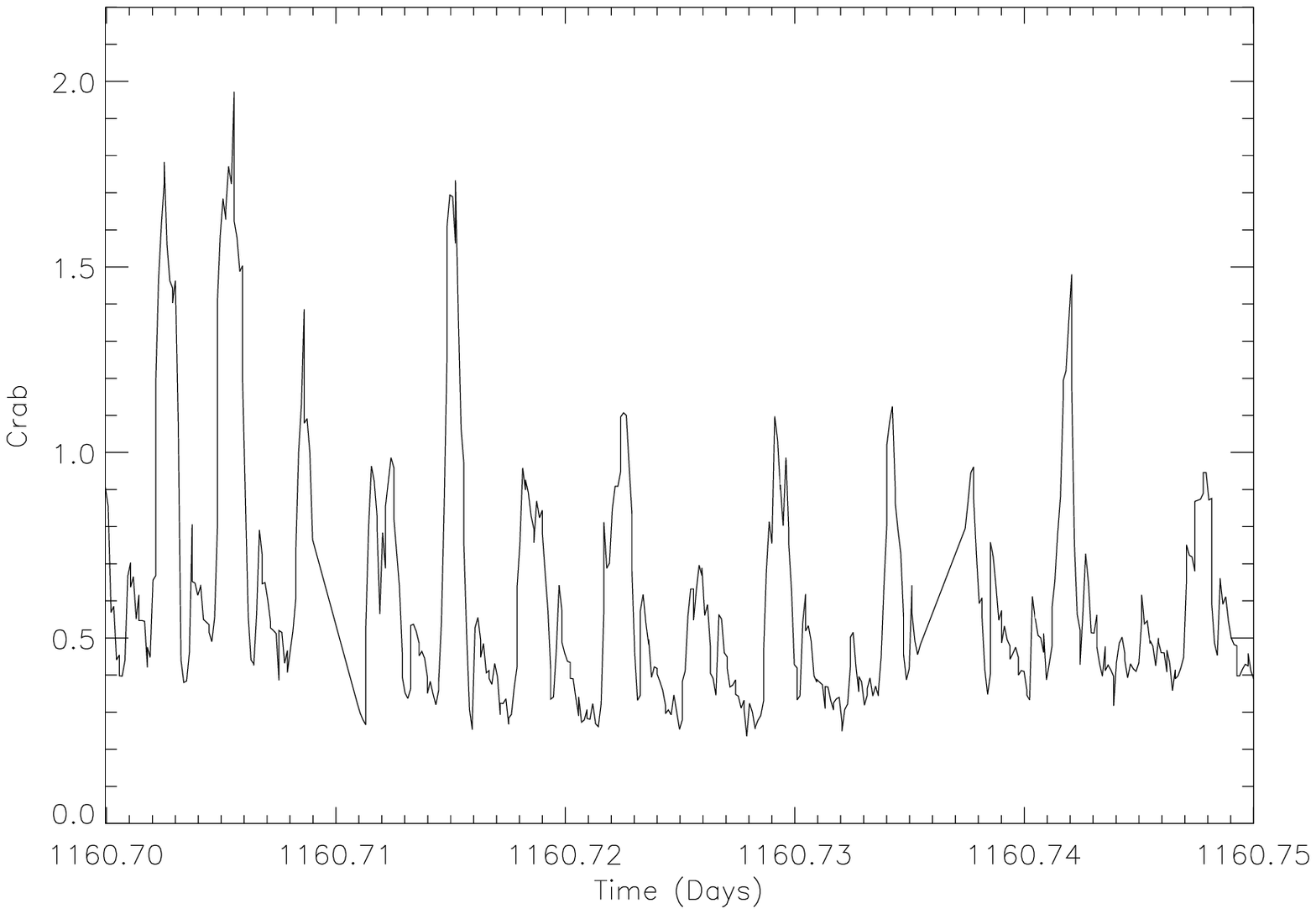} \\
 \includegraphics[angle=0,width=8cm]{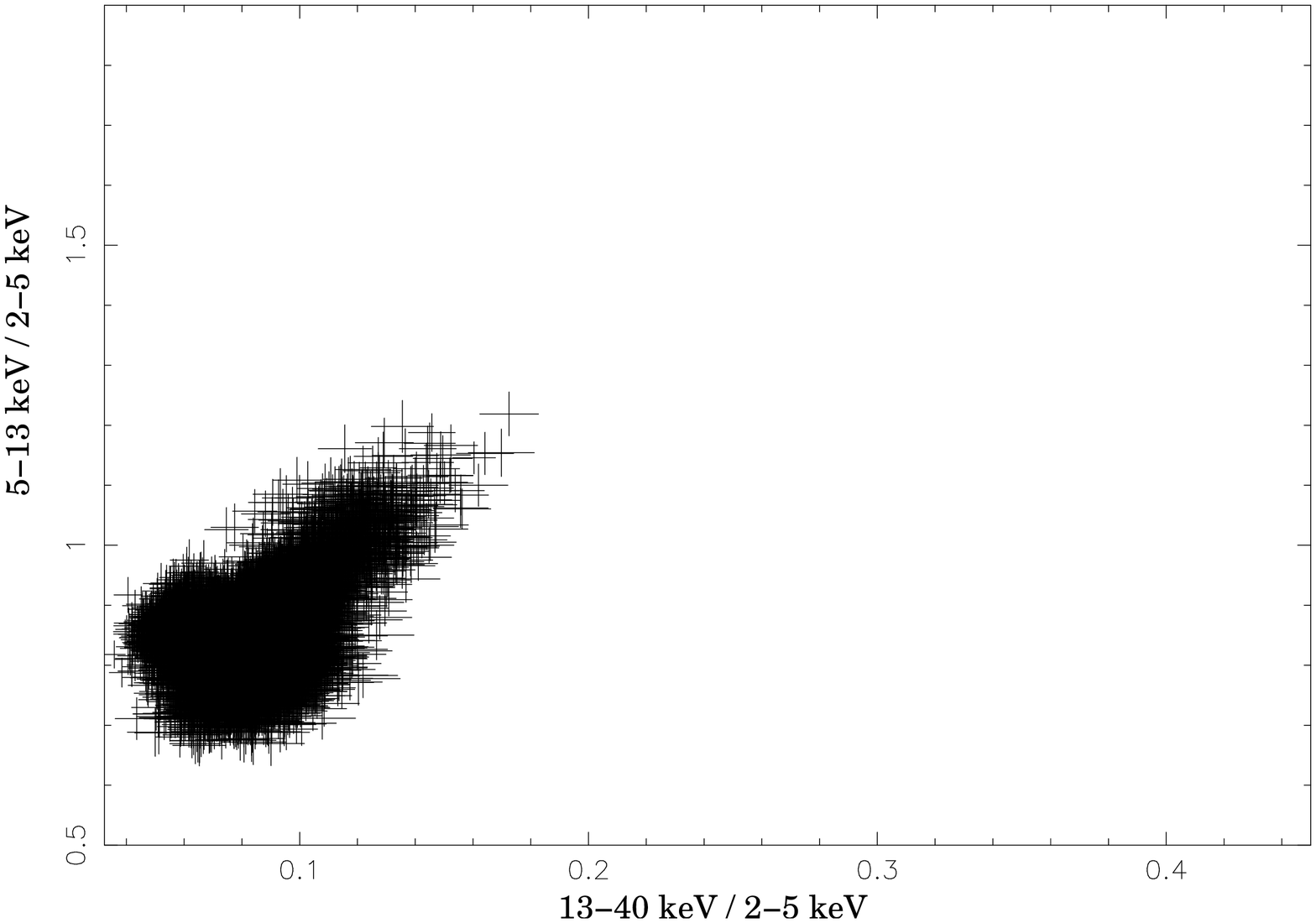} \\
\end{tabular}
\caption{The JEM~X-2 lightcurves. {\bf Top:} the whole 100~ks observation showing 
  strong variability. {\bf Middle:} a zoom covering 1.2~hours. The bin size
  is 8~s. {\bf Bottom} RXTE/PCA color-color diagram. The time bin  is 1~s. The
scale allows for a direct comparison with Figs. 2a and 2b in Belloni et al. (2000)}\label{jemx}
\end{figure}

The Joint European X-ray monitor, JEM-X (Lund et al. 2003), consists
  of two identical coded mask 
  instruments designed for X-ray imaging in the range of 3--35 keV with an angular 
  resolution of 3 arcmin and a timing accuracy of 122~$\mu$-s.
At present, only JEM~X-2 is operational, so all results shown here are from 
  that instrument.
The top left-hand side of Figure~\ref{jemx} shows the JEM~X-2 
  lightcurve from the whole 100~ks observation.
As can be seen, the source was in a very variable state, with the flux varying 
  between 0.25 to 2 Crab with a mean of $\sim0.5$~Crab.
The other panel shows a zoom of the lightcurve, which shows 
  particularly striking rapid oscillations.
Of particular interest are the main peaks separated by $\sim 5$
minutes. Although this kind of variability  resembles  the 
$\rho$-heartbeat, $\nu$ and $\phi$ oscillations (variability classes
  of Belloni et al. (2000)), these are more uniform
 and occur on shorter timescales. 
Our RXTE observations did not cover the entire
100 ks INTEGRAL data but  were simultaneous. They confirm the variability seen by JEM~X-2. 
We  produced a color-color (CC) diagram for the RXTE data 
in the same manner as in Belloni et al. (2000) (note that the energy-channel 
conversion for PCA corresponds to epoch 5). The resultant plot is shown in Fig.~\ref{jemx}.
This CC-diagram and the JEM~X-2 lightcurve seem to indicate a new
  type of 
variability not seen in the 12 classes by Belloni et al. (2000). 
The JEM~X-2 lightcurves were also built in four energy bands and a  
 CC-diagram ( (6--15 keV)/(3--6 keV)  vs (15--35 keV)/(3--6 keV) ) 
 constructed for the mean pulse profile of the 5-min oscillations. 
The resulting CC-diagram was very similar to that of RXTE shown in
  Fig. 5.  
The `ups' were harder than the `downs' -- this behavior is opposite  
to that seen in the $\rho$, $\nu$ and $\phi$ oscillations.

\begin{figure}[h]
 \includegraphics[angle=0,width=8.5cm]{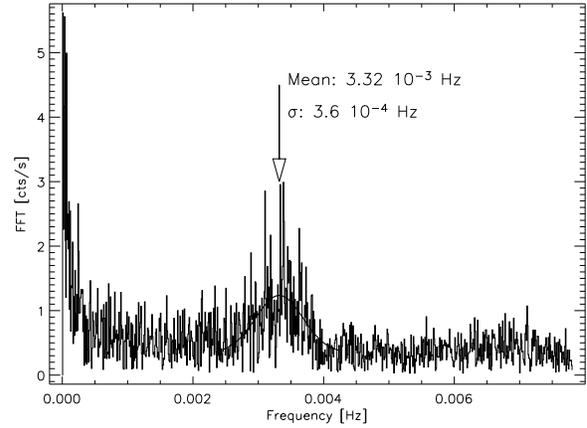} 
\caption{The JEM~X-2 power spectrum.}
  \label{pds}
\end{figure}

In order to have a first overview of the JEM~X-2 capability in timing
   analysis, we produced a simple Fourier Transform of the whole JEM~X-2 
   lightcurve (Fig.~\ref{pds}).
After removing artefacts due to some gaps in the data, we can identify
   a Quasi-Periodic Oscillation (QPO) at a frequency 
   $3.32\times10^{-3}$ Hz ($\sigma=3.6\times10^{-4}$ Hz). 
At the current stage, the exact power of the feature cannot be known
   exactly since it requires a precise background estimate.
The observation of such a low QPO with JEM~X-2 is, however, of prime
   importance since it opens the studies of low
   frequencies (difficult to access with RXTE due to its 90-minute
   orbit, and usually shorter observations) through the long,
   continuous INTEGRAL observations. The frequency of this  
QPO is in agreement with the 5-minute timescale of the main peaks in the 
lightcurve. The fact that such a feature is detected from the Fourier transform
of the whole 100 ks JEM~X-2 lightcurve indicates that this class of variability
is dominated by the  5-min  oscillations.
A deeper analysis of the variability is in progress and will be
   published elsewhere.

\subsection{Preliminary spectral analysis}

   \begin{figure}[h]
   \centering
   \includegraphics[angle=-90,width=8cm]{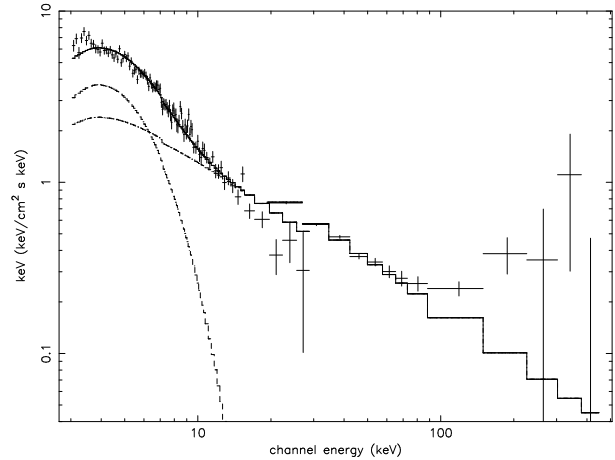}
   \caption{The JEM~X-2 (3--20~keV) and ISGRI (20--450~keV) spectrum. 
    The JEM~X-2 spectrum is a mean from $\sim37$ minutes, while 
    the ISGRI spectrum is accumulated over the whole observation
   of 100~ks. The residuals seen above $\sim$~300~keV are 
    probably due to the background.
    No systematics have been included, and the RMF and ARF
   in the case of ISGRI are preliminary matrices. 
   Calibration is still in progress.}
              \label{spectrum}
    \end{figure}

ISGRI spectra were extracted with a pre-flight 128 channels redistribution
  matrix file (rmf) and despite source variability, all spectra from
  all pointings were averaged. 
Figure~\ref{spectrum} shows the ISGRI spectrum accumulated over the
  whole observation of $\sim$100~ks.
Note that we used an ancillary response file (arf) preliminarily 
   corrected compared to the pre-flight matrix (P. Laurent,
  private communication).
JEM~X-2 spectra from science window 26, i.e. 
   the interval covering (Fig.~\ref{jemx}) IJD=1160.79063 to
  1160.81615, were extracted.
Data above 450~keV in the case of ISGRI and above 30~keV in the case
  of JEM~X-2 were ignored, and some bins above 100~keV were
  co-added.

The XSPEC package v11.2.0  was used for the fitting. 
The resulting co-added spectrum and model is shown in Fig.~\ref{spectrum}.
The model used was photoelectric absorption, plus a disk blackbody and
   powerlaw. 
N$_{\rm H}$ was frozen to 5$\times10^{22}$ (Lee et al. 2002). 
The fit yielded an inner temperature of the disk of 1.1$\pm$0.17~keV
  and a photon index 2.97$\pm$0.3. 
This photon index is what is  expected  for the medium
luminosity level (see Fig. 1) during which the present observations 
were performed (e.g. Muno et al. 1999, Vilhu et al. 2001).
The normalizations for both ISGRI and JEM~X-2 were left free.
The best fit led to a normalization of $13 \pm 1$ 
photons/keV/cm$^2$/s at 1 keV for JEM~X-2, and 
$16\pm 2$ photons/keV/cm$^2$/s for ISGRI. The 2--20 keV unabsorbed flux was 
$\sim 2\times10^{-8}$ erg~cm$^{-2}$~s$^{-1}$. 
The blackbody component accounts for 51$\%$, 
of the 2--20 keV unabsorbed flux.

The flux in the 20--450~keV range is $1.39\times10^{-9}$~erg~cm$^{-2}$~s$^{-1}$,
  which corresponds to a luminosity of $\sim 2.1\times10^{37}$~erg~s$^{-1}$
  for a distance of 11.5~kpc.
However, it must be noted that the source was in a very variable state
  and is known to vary on all timescales, so the resulting spectrum is
  in fact an average over the variability.




\section{Summary}

Although major analysis of the Open Time observations of GRS~1915$+$105 
  is being undertaken and the results will be presented elsewhere, here
  we want to stress the following: \\


\noindent $\bullet$ GRS~1915$+105$ was clearly detected by all 
          three of INTEGRAL's X-ray and gamma-ray instruments 

\noindent $\bullet$ The source was highly variable as demonstrated by both
          the JEM~X-2 3--35~keV and ISGRI 20--40 and 40--80~keV
          lightcurves. This class of variability seems to be new in comparison
          with previous studies.

\noindent $\bullet$ The source brightness varied from 0.25 to 2 Crab in the 
          JEM~X-2 energy range and between $\sim30-100$~mCrab
          ($\sim24-50$~mCrab) in the 20--40~keV (40--80~keV) ISGRI ranges 

\noindent $\bullet$ Assuming a distance of 11.2~kpc, 
          GRS~1915$+$105 had an averaged 20--450~keV luminosity
          $\sim 2.1\times10^{37}$~erg~s$^{-1}$

\begin{acknowledgements}
DCH and OV acknowledge the
Academy of Finland, TEKES, and the Finnish space research program
ANTARES for financial support in this research.
AAZ has been supported by KBN grants 5P03D00821, 2P03C00619p1,2
and PBZ-054/P03/2001, and a grant from FNP.
The authors acknowledge P. Laurent, M. Cadolle-Bel for kindly
providing preliminary ISGRI response matrices, and V. Beckmann for
useful help with the SPI data analysis. JR acknowledges financial support from
the French space agency (CNES).
We acknowledge quick-look results provided by the ASM/RXTE team.
We thank the referee for valuable comments which made this paper
more complete.

\end{acknowledgements}

\bibliographystyle{aa}

\end{document}